\documentclass[conference]{IEEEtran}

\usepackage{amsmath}
\usepackage{comment}
\usepackage{amsfonts}
\usepackage{enumitem}
\usepackage{tablefootnote}
\usepackage{cite}
\usepackage{xcolor}
\usepackage{algorithm} 
\usepackage{algpseudocode} 
\usepackage{bm}
\usepackage{threeparttable, tablefootnote}

\makeatletter
\newcommand{\algmargin}{\the\ALG@thistlm}   
\makeatother
\algnewcommand{\parState}[1]{\State%
    \parbox[t]{\dimexpr\linewidth-\algmargin}{\strut #1\strut}}

\ifCLASSINFOpdf
\usepackage[pdftex]{graphicx}
\else
\fi

\ifCLASSOPTIONcompsoc
  \usepackage[caption=false,font=normalsize,labelfont=sf,textfont=sf]{subfig}
\else
  \usepackage[caption=false,font=footnotesize]{subfig}
\fi

\newcounter{myfootertablecounter}

\usepackage{soul}

\begin{document}

\title{On Learning Intrinsic Rewards for Faster Multi-Agent Reinforcement Learning based MAC Protocol Design in 6G Wireless Networks}% }
%\title{Faster MAC Protocol Learning via Intrinsic Reward Stimulation for 6G Wireless Networks}% }
%\title {Fast MAC protocol Learning in 6G wireless networks trough lifetime based intrinsic reward stimulation}% }

\author{
    \IEEEauthorblockN{
        Luciano Miuccio\IEEEauthorrefmark{1}, Salvatore Riolo\IEEEauthorrefmark{1}, Mehdi Bennis\IEEEauthorrefmark{2}, and  Daniela Panno\IEEEauthorrefmark{1}
    }
    \IEEEauthorblockA{\IEEEauthorrefmark{1} Department of Electrical, Electronics and Computer Engineering, University of Catania, Italy}
    \IEEEauthorblockA{\IEEEauthorrefmark{2} Centre for Wireless Communications, University of Oulu, Finland}
	emails: luciano.miuccio@phd.unict.it, \{salvatore.riolo, daniela.panno\}@unict.it, mehdi.bennis@oulu.fi
}

%\author{\IEEEauthorblockN{Luciano Miuccio, Daniela Panno, Salvatore Riolo}
%	\IEEEauthorblockA{Dept. of Electrical, Electronics and Computer Engineering, University of Catania, Italy\\
%	%	University of Catania
%	%	Catania, Italy\\
%		Email: luciano.miuccio@phd.unict.it, daniela.panno@dieei.unict.it, salvatore.riolo@unict.it}}
%\author{\IEEEauthorblockN{}
%		\IEEEauthorblockA{
			%	University of Catania
			%	Catania, Italy\\
%				Emails: luciano.miuccio@phd.unict.it, salvatore.riolo@unict.it,  \{sumudu.samarakoon, mehdi.bennis}@oulu.fi,\}@oulu.fi}

\maketitle

\begin{abstract}

In this paper, we propose a novel framework for designing a fast convergent multi-agent reinforcement learning (MARL)-based medium access control (MAC) protocol operating in a single cell scenario. 
The user equipments (UEs) are cast as learning agents that need to learn a proper signaling policy to coordinate the transmission of protocol data units (PDUs) to the base station (BS) over shared radio resources. 
%Due to the partial observability of the global states, i
In many MARL tasks, the conventional centralized training with decentralized execution (CTDE) is adopted, where each agent receives the same global extrinsic reward from the environment. However, this approach involves a long training time. To overcome this drawback, we adopt the concept of learning a per-agent intrinsic reward, in which each agent learns a different  intrinsic reward signal based solely on its individual behavior. Moreover, 
in order to provide an intrinsic reward function that takes into account the long-term training history, we represent it as a long short-term memory (LSTM) network.
%in order to consider a long-term training history, the intrinsic reward function is represented as a long short-term memory (LSTM) network. 
As a result, 
%each agent has two modules, i.e., a policy network and an intrinsic reward network. 
each agent updates its policy network considering both the extrinsic reward, which characterizes the cooperative task, and the intrinsic reward that reflects local dynamics.  
%to adapt the decision-making while the intrinsic reward network is a long short-term memory (LSTM) network, updated with a periodicity of several episodes (i.e., a lifetime), which differently stimulates the agent at each time step on the basis of its lifetime history. 
%updates its policy network considering both the extrinsic reward, which characterizes the cooperative task, and the intrinsic reward, which diversifies the individual behavior.
%The intrinsic reward function is represented as a long short-term memory (LSTM) network
%The policy network is updated for each episode to adapt the decision-making while the intrinsic reward network is , updated with a periodicity of several episodes (i.e., a lifetime)
%learning the proper MAC protocols in
%this systems consisting of multiple agents acting and learning in the same shared environment takes a long training time
%To this end, the new framework combines concepts of intrinsic reward learning and a lifetime update. Each UE has two modules, i.e., a policy network and an intrinsic reward network. The policy network is updated for each episode to adapt the decision-making while the intrinsic reward network is a long short-term memory (LSTM) network, updated with a periodicity of several episodes (i.e., a lifetime), which differently stimulates the agent at each time step on the basis of its lifetime history. 
The proposed learning framework yields a faster convergence and higher transmission performance compared to the baselines.
Simulation results show that
%, when the scenario is too simple, the introduction of the new features do not lead to any improvement. Conversely, in the case of a challenging configuration, 
the proposed learning solution yields 75\% improvement in convergence speed
compared to the most performing baseline.
\end{abstract}

% Note that keywords are not normally used for conferences.

\IEEEpeerreviewmaketitle

\section{Introduction} 

The emergence of data-driven medium access control (MAC) protocols can provide a cost-effective, flexible approach to boost the performance of beyond 5G (B5G) and 6G networks. %that are envisioned to support heterogeneous services and applications \cite{8412482}.
%However, to provide these protocols with reduced time, effort, and cost is still a challenge. In particular, designing a MAC protocol that permits to be deployed timely  is a major challenge, especially when the scenario becomes more complex.
To address this problem, %machine learning (ML) can be used \cite{9247527}, and in particular, 
multi-agent reinforcement learning (MARL) methods enable agents to learn an optimal policy by interacting in the same environment \cite{art_marl1}. 
%However, the multi-agent environments suffer of several issues. First,  the agents need to learn their policies concurrently. The action of an agent may affect the rewards of other agents, and the state transition is decided by the joint actions of all the agents. 
%To ensure the convergence of the learning algorithm, the learning agent should account for the actions of other agents when making action decisions. Second, the learning agents are typically unable to have a full observation on the environments such as large-scale communication networks in reality. They usually make autonomous action decisions based on their partial observations. 
%Since the agents cannot access the actions of other agents, they may not solve the non-stationarity issue. Therefore, the convergence of the policy of the learning agent cannot be guaranteed. 
%To solve the above issues, MARL enables the agents to use various state/observation structures, which result in different learning schemes, such as, fully centralized learning, decentralized learning with networked agents, independent learning (IL), and centralized training with decentralized execution (CTDE). 
Current works,  such as \cite{nokia_paper_1, globecom}, have studied the MAC protocol learning in a single cell scenario, where user equipments (UEs) need to deliver MAC protocol data units (PDUs) to the base station (BS) sharing the same radio channel. UEs are cast as reinforcement learning (RL) agents  that are trained to learn a new MAC protocol from their partial observations of the global state. 

However, despite the good performances shown at the end of the training procedure, learning efficient and robust MAC protocols consisting of multiple agents acting and learning in the same shared environment requires very long training time. This aspect prevents the applicability of this approach to a dynamic wireless environment that requires retraining to adapt the MAC protocol to changing environments. 
%tois a major challenge, especially when the scenario becomes more complex.
The main causes of slowness in training stem from the partial observability and  non-stationarity of the MARL problem (i.e., transitions from a state to another depend on the actions of all agents) \cite{arxiv.1906.04737}.
In addition to this,  \cite{nokia_paper_1, globecom} are based on the conventional centralized training and decentralized execution (CTDE) paradigm and parameter sharing technique \cite{param_sharing1} that further slows down the convergence time. 
%First, the use of the conventional centralized training and decentralized execution (CTDE) paradigm. Second, the adoption of the parameter sharing technique \cite{param_sharing1}. 

Specifically, on the one hand, CTDE allows agents to learn their local policies in a centralized way while retaining the ability of decentralized execution. During the training phase, 
the environment assigns the same global reward to all agents
without distinguishing their own contributions. 
As a consequence, only a subset of agents contributes to the reward, and so, during training an agent may be punished even if it has acted optimally, or rewarded even if it has acted wrongly. Clearly, this approach induces slow  and unstable policy learning.
%the agents only have idea of the collective reward signal thus hindering the policy optimization.
 %The CTDE paradigm considers for a 
 
On the other hand, the parameter sharing technique consists in simultaneously learning a single shared policy for multiple agents, which boosts scalability.
% since it creates a common policy that can be transferred among several UEs. 
However, UEs may compete at the same time to transmit their own packets in the same shared channel. In other words, even if two UEs have the same observation, the action taken by one UE should be different from the action taken by the other one to avoid interference. Updating the same policy for every agent can create adverse actions that slow down convergence during learning. 

In light of the above considerations, the main contribution of this paper is to propose a \textbf{novel framework that provides faster convergence to the MAC protocol learning problem}. Specifically, we consider the same communication scenario studied in \cite{nokia_paper_1, globecom}
 and introduce our innovative learning framework with the following features. 
 
 First, we adopt an enhanced version of the CTDE paradigm, which in addition to the global reward signal, leverages  for each agent, a different local intrinsic reward signal based on its individual behavior. % considering its long-term training history. 
 This idea is inspired by intrinsic reward learning introduced in \cite{int_ext_reward,intrinsic_reward} for a single-agent environment.  Different from the global reward given by the environment  (termed as “extrinsic reward”) that is hand-designed, the intrinsic reward is automatically learned by each agent. 
 
 Second, the proposed solution avoids the use of the parameter sharing technique and, instead, considers that each agent has two independent modules, namely, a policy network and an intrinsic reward network. The policy network learns the optimal policy per agent, while the intrinsic reward network provides additional reward signal to the policy network. 
 
 %Third, it designs an efficient updating routine for the agent's modules that considers two different periodicities, i.e., the episode for the policy network and a sequence of several episodes (called lifetime) for the intrinsic reward network. This feature is inspired by \cite{intrinsic_reward}.
 
 Simulation results show that in complex scenarios and adopting the multi-agent proximal policy optimization (MAPPO) algorithm, %\cite{MAPPO_1},  
 the proposed learning framework yields a $75\%$ improvement in convergence speed, and about $4\%$  improvement in transmission performance compared to conventional CTDE without parameter sharing technique, and even better results with respect to a baseline consisting of both CTDE and parameter sharing.
 
The rest of the article is organized as follows. 
The system model and the formalization of the cooperative MARL problem are described in Section \ref{system}. The proposed approach is detailed in Section \ref{approach}. Finally, the numerical simulation results and conclusions are drawn in Section \ref{sim_and_ana} and Section \ref{conclusion}, respectively.

\section {System Model and MARL Formulation}
\label{system}
Consider a single BS serving a set $\mathcal{N}$ of $N$ homogeneous UEs needing to deliver $P$ MAC PDUs to the BS. 
%Both the data plane and the control plane are considered. In the remainder of this paper, we denote a data PDU transmitted in the data plane with “dPDU”, and a signaling PDU transmitted in the control plane with “sPDU”.
The network nodes exchange control messages encapsulated inside signaling PDUs (sPDUs) through the downlink (DL) and uplink (UL) control channels, which are assumed to be dedicated and error free.
%, so without any contention or collision. 
As regards data transmission, UEs send data PDUs (dPDUs) using the same physical uplink shared channel (PUSCH) operating according to a time division multiple access (TDMA) scheme, which leads to possible collisions. Specifically, for each time step $t$ a UE can send one dPDU, and this dPDU is successfully received by the BS only if a single UE out of $N$ has transmitted it. 

\textbf{Control plane}: let $\mathcal{M}_{\text{UE,s}}=\{0,1\}$ be the set of possible messages sent by the UEs, and $\mathcal{M}_{\text{BS}}=\{0,1,2\}$ be the set of DL messages. At each time step $t$, the BS can send to each UE $i\in\mathcal{N}$ only one message $m_i^t\in\mathcal{M}_{\text{BS}}$, and each UE $i$ can send one signaling message $a_{i,\text{s}}^t\in\mathcal{M}_{\text{UE,s}}$ to the BS. Specifically, $m_i^t=2$ represents an acknowledgement (ACK) message that confirms that a dPDU sent from UE $i$ has been correctly received at the BS in the previous time step $t-1$, $m_i^t=1$ refers to a scheduling grant message to UE $i$, and $m_i^t=0$  to indicate that no access is granted for  UE $i$. As for UEs, $a_{i,\text{s}}^t=1$ means sending an access request to reserve time step $t+1$ for  transmission, while $a_{i,\text{s}}^t=0$ means do not transmit any signaling message.

\textbf{Data plane}: each UE $i$ has a dPDUs storage capability, modeled as a buffer with first-in first-out (FIFO) policy, which contains at most $P$ dPDUs. We denote with  $b_i^t\in\mathcal{B}=\{0,1,\dots,P\}$ the buffer status at time $t$, and we assume that the buffer starts full. For each time step $t$, UE $i$ is able to transmit a dPDU or to delete it. Specifically, the data plane action is denoted as $a_{i,\text{u}}\in \mathcal{M}_{\text{UE,u}}=       \{0,1,2\}$, where $a_{i,\text{u}} = 1$ means that the UE transmits the first dPDU in its buffer (if any), $a_{i,\text{u}} = 2$ means it deletes the first dPDU in the buffer, and $a_{i,\text{u}} = 0$ to do nothing. 

We assume that the BS is a MAC expert agent, i.e., it adopts a MAC protocol that is not learned. In detail, at each time step $t$, if the BS receives more scheduling requests from the UEs, then it chooses one of the requesting UEs at random and a scheduling grant is sent in response. If the UE has made a successful data transmission concurrently with the scheduling request, then the BS will ignore this scheduling request and send only an ACK message to it.
%The PUSCH is modeled as a packet erasure channel, where a PDU is incorrectly received with a fixed probability equal to the transport block error rate (TBLER). 
%\st{The set of UEs is denoted as $\mathcal{N}$.}

%We suppose that each time step $t$ is dimensioned so that each UE in $\mathcal{N}$ can transmit exactly one dPDU inside a time step, including the propagation delay within the medium and other additional delays. 
%In the following, we denote with $x^t$ the generic variable $x$ at time step $t$. 
%\st{Otherwise, if more UEs have transmitted} 
%\sps{If multiple UEs simultaneously transmit} 
%their dPDUs, a collision occurs and the BS cannot correctly decode the received dPDUs. 
% \spsc{X}{Change $\mathcal{X}_{UE}$ to $\mathcal{X}_{\text{UE}}$ since the subscript is not referring to a variable. Otherwise, it could be interpretted as a product of $U$ and $E$ or a tuple defined over the entities $U$ and $E$. This is valid for all other notations throughout the paper.}

\subsection{Multi-Agent Reinforcement Learning Formulation}
The goal is to find the optimal MAC protocol adopted by UEs that maximizes the number of unique dPDUs successfully received by the BS, while minimizing the time spent to do so.
% \begin{equation}
%     \max \left\{\frac{N_{\text{p,rec}}}{T_{\text{tr}}}\right\},
% \end{equation}
% where $N_{\text{p,rec}}$ represents the number of unique dPDUs received by the BS, and $T_{\text{tr}}$ the time spent to receive $N_{\text{p,rec}}$ packets.
To effectively reach this goal, we propose to cast the UEs as MAC learning agents and the protocol learning problem as a cooperative and multi-agent partially observable Markov decision process (MPOMDP). The system can be described as a tuple as  $\langle$$\mathcal{N}$, $\mathcal{A}$, $\mathcal{S}$, $\mathcal{O}$, $\pi_{i}$, $R_{\text{ext}}$,$\gamma$$\rangle$.
Let $\mathcal{N}$ denote the set of $N$ homogeneous learning agents (i.e., UEs). Each agent $i\in\mathcal{N}$ at time step $t$ has a partial observation of the global state defined as $o_i^t=(b_i^t,b_i^{t-1},a_i^{t-1},m_i^{t-1},\dots,b_i^{t-M},a_i^{t-M},m_i^{t-M})$, where $M$ is the memory length.  Accordingly, let ${a}_i^t=(a_{i,\text{u}}^t,a_{i,\text{s}}^t)$ indicate the action taken by agent $i$, which involves both data and control plane. Each agent $i$ shares the same observation and action space, denoted as $\mathcal{O}$ and $\mathcal{A}$, respectively. Clearly, $\mathcal{A}= \mathcal{M}_{\text{UE,u}} \times \mathcal{M}_{\text{UE,s}} = \{A_1, \dots, A_{|\mathcal{A}|}\}$. Let $\pi_{i}\left(a_i^t\mid o_i^t\right)\colon\mathcal{O}\times\mathcal{A}\rightarrow[0,1]$ be a stochastic policy for agent  $i$, that is, the probability of choosing a given action $a_i$ given that agent $i$ is observing $o_i$.
%, and $l_i^t=\log\pi_{i}\left(a_i^t\mid o_i^t\right)$.
For sake of clarity, we also introduce  $\bm{o}^t=[o_1^t,\dots,o_N^t]$, $\bm{a}^t=[a_1^t,\dots,a_N^t]$, and $\bm{\pi}=[\pi_1,\dots,\pi_N]$.
%, and $\bm{l}^t=[l_1^t,\dots,l_N^t]$. %$\pi=\{{{\pi}}_i \}^N_{i = 1}$.

At time step $t$, each agent $i$ observes $o_i^t$ and selects an action $a_i^t$ according to its own policy $\pi_{i}$. At time step $t+1$, in conventional MPOMDP, each agent receives from the environment an extrinsic reward $R_{\text{ext}}^{t+1}$, which is the same for all agents and quantifies the benefit of the joint actions performed by all the $N$ agents. This design decision reflects the objective of optimizing the performance of the whole network, rather than that of individual agents.
We define an episode as a finite sequence of agent-environment interactions lasting $T_\text{ep}$ time steps. For each episode, we define the episodic cumulative extrinsic return as
\begin{equation}
G_{\text{ep},\text{ext}} = \sum_{t=0}^{T_{\text{ep}}-1}\gamma^t R_{\text{ext}}^{t+1},
\end{equation}
where $\gamma$ is a discount factor. Since maximizing $G_{\text{ep},\text{ext}}$  represents the goal of the reinforcement learning problem, the values of $R_{\text{ext}}^{t+1}$ in each time step $t$ should be properly designed. Here, we leverage the simple approach \cite{nokia_paper_1}, where $R_{\text{ext}}^{t+1}\in\{-1,0\}$, with $R_{\text{ext}}^{t+1}$ equal to 0 only if a dPDU has been received correctly at time step $t$ or if all dPDUs sent by each UE have been already received correctly in the previous time steps. This means that $G_{\text{ep},\text{ext}}$ reaches its maximum value (i.e., 0) when all packets have been received at minimum time, and the minimum value of $G_{\text{ep},\text{ext}}$ is assumed when no packets have been received correctly. We emphasize that the  selection of extrinsic reward functions are typically hand-designed.
%, and the designer might provide better extrinsic reward functions to the same MARL problem to speed up the convergence time. 
However, finding a good reward function is not straightforward and  requires a high expertise and domain knowledge of the designer. 
%or this operation can be done via time-consuming exhaustive searches. 
Moreover, the extrinsic reward is strongly goal  or task-specific, which
%related to the goal, and so, this  time-consuming operation 
limits its applicability to other use cases and goals.
Let $J_{\text{ep},\text{ext}}(\bm{\pi})$ denote the expected episodic cumulative extrinsic return obtained when each agent $i$ follows its own policy $\pi_i\in\bm\pi$, i.e.,
\begin{equation}
    J_{\text{ep},\text{ext}}(\bm{\pi})= \mathbb{E}_{\textbf{o}^0,\textbf{a}^0,\dots\textbf{o}^{T_\text{ep}-1},\textbf{a}^{T_\text{ep}-1}} \left [G_{\text{ep},\text{ext}}  \right ],
\end{equation}
where $a_i^t\sim\pi_{i}\left(a_i^t\mid o_i^t\right), \forall i \in \mathcal{N}$.
The objective of the MARL problem is to find optimal policies $\bm{\pi}^*$ that maximize $J_{\text{ep},\text{ext}}$. For doing this, we adopt the CTDE paradigm. Several MARL techniques can be used, ranging from simple value-based approaches (e.g., tabular Q-learning \cite{sutton2018reinforcement}) to on-policy algorithms (e.g., MAPPO \cite{MAPPO_1}). In general, all approaches consider 
%for each agent $i$ 
an independent policy parameterized by ${\theta_i}$ for each agent $i$ and denoted as $\pi_{\theta_i}$. Each agent updates independently its own parameter $\theta_i$ by maximizing the expected extrinsic reward. In addition to this, thanks to the homogeneous nature of UEs and the use of the same cumulative extrinsic reward $J_{\text{ep},\text{ext}}$, another possible approach is to learn a shared optimal policy $\bm{\pi}^*$
by leveraging the concept of parameter sharing \cite{param_sharing1}. %\cite{param_sharing1, param_sharing2}.
%In this paper, we assume that the policy parameters are updated at each time step $t$, but in general they can be updated less frequently. 

\section{Proposed approach}
\label{approach}
In this section, we formally present our approach that aims to automatically speed up the convergence time by adopting both the concept of extrinsic reward empowered by an intrinsic reward \cite{int_ext_reward} and the concept of lifetime \cite{intrinsic_reward}. Specifically, in \cite{int_ext_reward} the authors investigated in the case of a single agent environments the advantages of using  an intrinsic reward function parameterized by $\eta$ in addition to the conventional extrinsic reward. Differently from the hand-designed extrinsic reward, the intrinsic reward function is automatically learned by each agent to improve its learning dynamics. In this case, both the policy and intrinsic reward parameters are learned within a single episode. Conversely, in \cite{intrinsic_reward}, the authors propose to learn an intrinsic reward over a lifetime consisting of $N_\text{ep}$ episodes, instead of a single episode, to take into account the system dynamics. 
%So, they defined lifetime as the finite sequence of agent-environment interactions during the $N_\text{ep}$ episodes. 
In detail, the policy parameter $\theta$ is still updated episode-by-episode by considering only the cumulative episodic intrinsic reward, while the intrinsic reward parameter $\eta$ is updated within every lifetime to maximize the cumulative extrinsic reward over an entire lifetime. 
%Clearly, the change on $\eta$ would influence the extrinsic value through the change in the policy parameters. 
% The main advantage of the adoption of the intrinsic reward is that it considerably simplifies the definition of the extrinsic reward. This advantage should not be seen as a one-time task. In fact, if in the same environment the task would change, the designer can define a simple extrinsic reward that follows the new task, while automatically the intrinsic reward will speed up the training time. 

Therefore, inspired by these works, we propose a new approach that incorporates the multi-agent intrinsic reward function in our system model. 
For the sake of clarity, we first define some terminologies. 
%\item \textbf{Lifetime:} A finite sequence of agent-environment interactions consisting of $N_\text{ep}$ episodes. 
\begin{itemize}
\item \textbf{Intrinsic reward function for agent $i$.} Defined as a function related to agent $i$ and parameterized by $\eta_i$. At the end of time step $t$, $R_{\text{in},\eta_i}^{t+1}$ is a scalar reward that takes into account the history of the entire lifetime of agent $i$ until time step $t$, including all its partial observations $([o_i^0,\dots,o_i^t])$, its selected actions $([a_i^0,\dots,a_i^t])$, and extrinsic reward values $[R_{\text{ext}}^{1},\dots,R_{\text{ext}}^{t}])$.
\item \textbf{Overall reward function for agent $i$.} Defined as a function related to agent $i$ made of two contributions. First, the extrinsic reward value $R_{\text{ext}}^{t+1}$ received from the environment, which is the same for all agents and quantifies the benefit of joint actions performed by $N$ agents. Second, the intrinsic reward value $R_{\text{in},\eta_i}^{t+1}$ that is learned independently by each agent $i$. For each agent $i$ and time step $t+1$,  the overall reward is given as
\begin{equation}
\label{overall_reward}
R_{\text{ov},i}^{t+1} = R_{\text{ext}}^{t+1} + \lambda R_{\text{in},\eta_i}^{t+1},
\end{equation}
where $\lambda\in[0,1]$ is a hyper-parameter that balances the weighted summation between the extrinsic reward and the intrinsic reward. 

\item \textbf{Episodic overall return.} For each episode $k$, we define the episodic overall return of agent $i$ as
\begin{equation}
\label{episodic_overal}
G_{\text{ep},\text{ov},i}^{(k)} = \sum_{t=0}^{T_{\text{ep}}-1}\gamma^t R_{\text{ov},i}^{t+1}.
\end{equation}
\item \textbf{Lifetime extrinsic return.} At the end of a lifetime, we define the lifetime extrinsic return as
\begin{equation}
\label{life_extrinsic}
G_{\text{life},\text{ext}} = \sum_{t=0}^{T-1}\gamma^t R_{\text{ext}}^{t+1},
\end{equation}
where $T$ is the number of steps per lifetime, i.e., $T=N_\text{ep}T_\text{ep}$. Using the lifetime return $G_{\text{life},\text{ext}}$ as the objective instead of the conventional episodic return $G_{\text{ep},\text{ext}}$ allows exploration across multiple episodes. 
\end{itemize}

\begin{figure}[!t]
	\centering
	\includegraphics[width=3.2in]{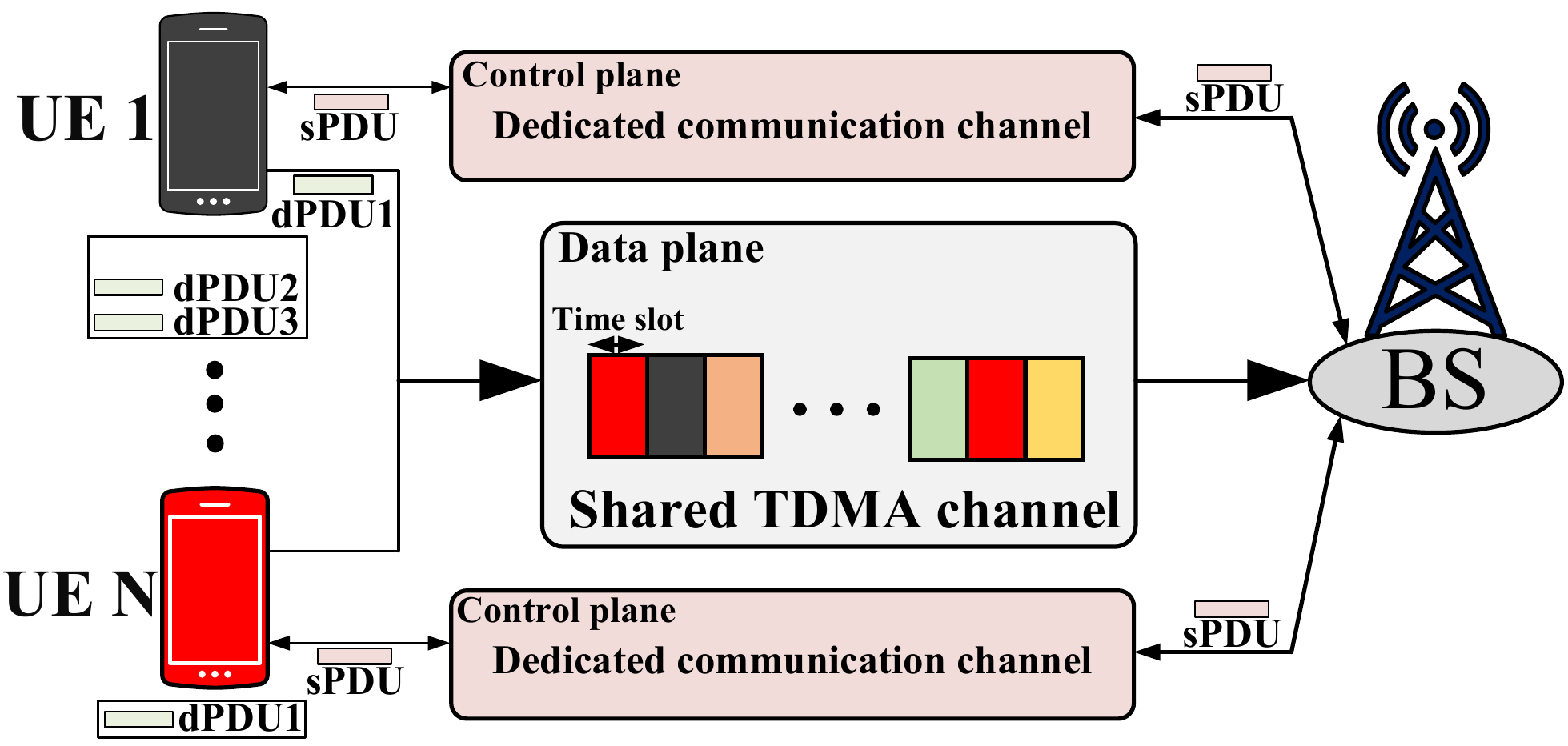}
	\caption{High-level depiction of the system model.}
	\label{system_model_fig}
\end{figure}

% To learn it, we adopt the CTDE paradigm. In general, several MARL techniques can be used, ranging from off-policy learning frameworks,
% such as MADDPG \cite{nokia_paper_2}, value-based approaches (e.g., tabular Q-learning \cite{nokia_paper_1}), to on-policy algorithms such as MAPPO \cite{MAPPO_1}. Among them, in this work, we adopt MAPPO since its on-policy nature is well-suited to the task of learning new MAC protocols.

%Specifically, in [nokia1] the authors resolve their MARL problem by using a tabular Q-learning approach, where  one central copy of two Q tables, which contain action value estimates for the data plane and the control plane, respectively, is updated regularly as experience from all MAC learners is collected. 

%In MADDPG, an actor-critic approach is adopted, where each agent has a decentralized actor network together with a common centralized critic. The actor network depends only on the agent’s partial observation in order to learn a decentralized policy, while the centralized critic receives all agents observations and actions to learn a joint action value function $Q(\textbf{o},\textbf{a})$.

\subsection{Architecture of the multi-agent framework}

\begin{figure}
    \centering
    \includegraphics[width=2.5in]{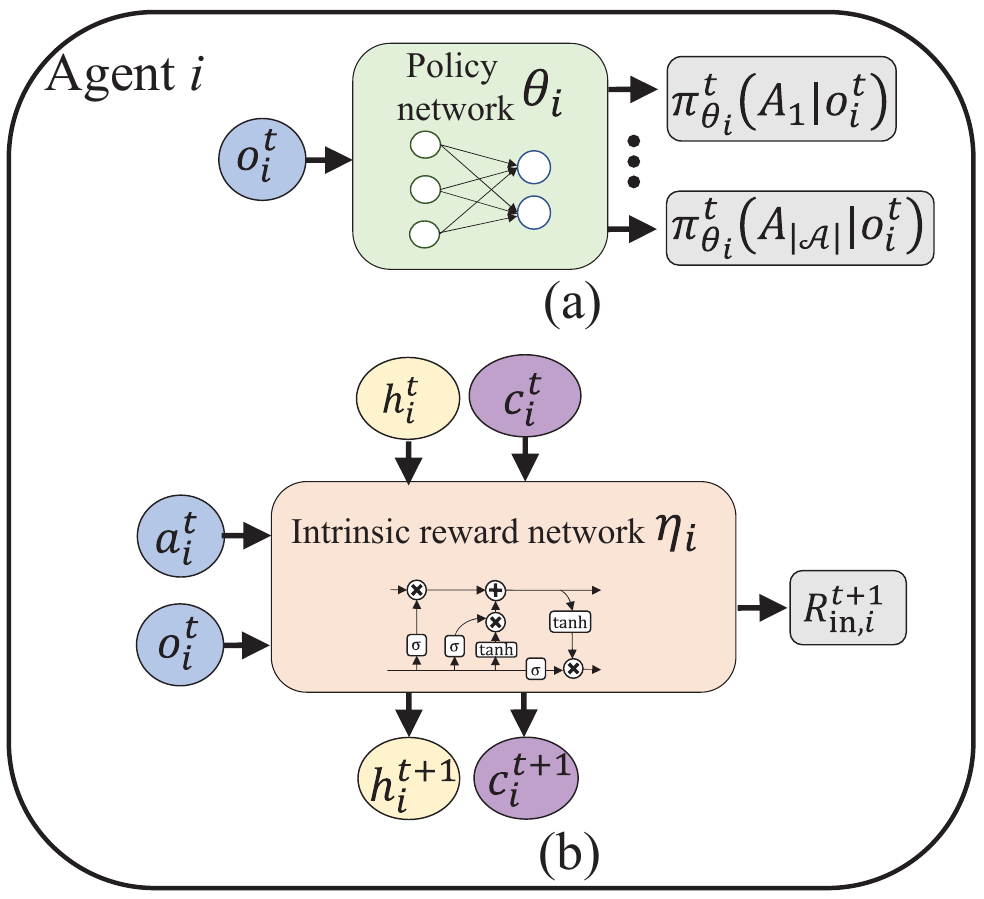}
    \caption{Learning networks inside each agent $i$. a) is the policy network parameterized by $\theta_i$ that, given the current observation $o_i^t$, outputs the probability of choosing the action $A_j$, for $j\in \{1, \dots, |\mathcal{A}|\}$. b) is the intrinsic reward network parameterized by $\eta_i$ that receives as input the current observation and the selected action ($o_i^t, a_i^t$) and is conditioned on the previous hidden state and cell state ($h_i^t, c_i^t$). It gives as output the intrinsic reward value $R_{\text{in},\eta_i}^{t+1}$ and generates the next tuple ($h_i^{t+1}, c_i^{t+1}$).}
    \label{policy_network}
\end{figure}

Each agent $i\in\mathcal{N}$ is equipped with two neural networks, as depicted in Fig. \ref{policy_network}. The first one represents the policy function $\pi_i$, and the second one represents the related intrinsic reward function.
The policy network (see Fig. \ref{policy_network}a) is a multi-layer perceptron (MLP) with weights $\theta_i$.
%that receives in input the current observation $o_i^t\in\mathcal{O}$ and gives an $|\mathcal{A}|$-dimensional output, where each dimension is the probability of choosing the action $A_j$, for $j\in \{1, \dots, |\mathcal{A}|\}$, given that the agent is observing $o_i^t$.
The intrinsic reward function is represented by a neural network providing as output a scalar reward that takes into account the long past history of agent $i$. For this reason, instead of adopting a conventional MLP, we exploit the characteristics of recurrent neural networks (RNNs). Unlike MLP, in RNNs the output from previous step is fed as input to the current step creating a feedback loop. As a consequence, the output provided at step $t$ takes into consideration not only the current input, but also what the network has learned from the previous inputs, involving internal memory capabilities. 
However, conventional RNNs are not able memorize data for long time and tend  to forget its previous inputs. To overcome this problem, we use an LSTM, which is a type of recurrent neural network that expands the memory capacity for long period of time \cite{LSTM_1}. The proposed LSTM is parameterized by $\eta_{i}$ for representing the intrinsic reward function, as shown in Fig. \ref{policy_network}b.
%For each agent $i$, the  is a recurrent neural network (RNN) parameterized by $\eta_{i}$. Since this function takes into account the history of the entire lifetime, we exploit the use of a Long short-term memory (LSTM) that permits to  maintain information in the memory for the long period of time \cite{LSTM_1}. 
%At each time step $t$, the network receives as input the tuple ($o_i^t, a_i^t$) and is also conditioned on the previous tuple ($h_i^t, c_i^t$), representing the hidden state and cell state, respectively. The network gives the intrinsic reward value $R_{\text{in},\eta_i}^{t+1}$ and generates the next tuple ($h_i^{t+1}, c_i^{t+1}$). At the end of an episode, both the hidden state and the cell state of the LSTM are preserved to the next episode. 

\subsection{Algorithm overview}
\begin{figure*}
    \centering
    \includegraphics[width=5in]{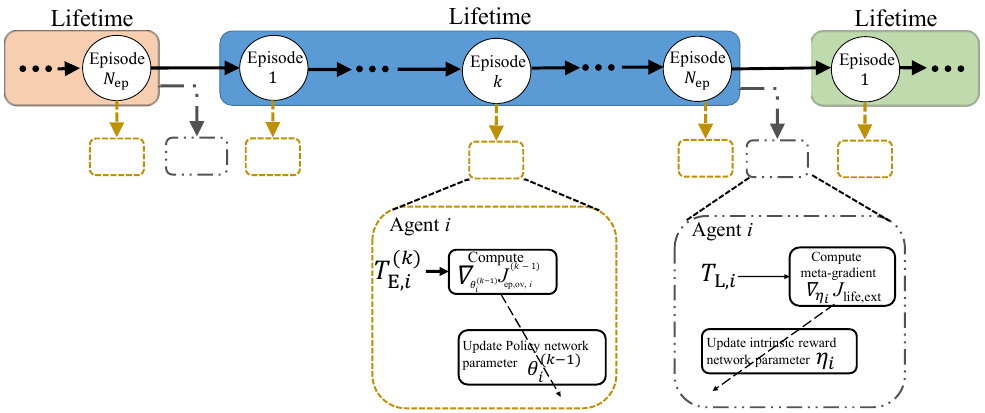}
    \caption{Training algorithm overview: for each episode $k$, each agent $i$ updates its policy parameter $\theta_i^{(k-1)}$ by exploiting the episode rollout $T_{\text{E},i}^{(k)}$ in the direction of the gradient of expected lifetime overall return $J_{\text{ep},\text{ov}, i}^{(k-1)}$. At the end of each lifetime, agent $i$ updates the intrinsic reward network parameter $\eta_i$ in the direction of the gradient of the expected lifetime extrinsic return $J_{\text{life},\text{ext}}$, by exploiting its lifetime rollout $T_{\text{L},i}$.}
    \label{training_periodicity}
\end{figure*}

A high level description of the proposed training algorithm related to each agent $i$ is presented in Pseudo-code \ref{pseudocode} and depicted in Fig. \ref{training_periodicity}. As shown, the updates of the policy network and the intrinsic rewards network are carried out with a different periodicity, corresponding to one episode and one lifetime, respectively. The periodicity of a lifetime permits to update the intrinsic reward network taking into account the long-term system dynamics.
%For the sake of clarity, we denote $k$ as the episode index inside a lifetime. 

\begin{algorithm} [H]                 % enter the algorithm environment
	\floatname{algorithm}{Pseudo-code}
	\caption{Network updates for agent $i$}          % give the algorithm a caption
	\label{pseudocode}                   % and a label for \ref{} commands later in the document
	{		\begin{algorithmic} [1]               % enter the algorithmic environment
			\renewcommand{\algorithmicrequire}{\textbf{Inputs:}} 
			\Require 
			learning parameters $\alpha$ and $\beta$, balancing parameter $\lambda$, discount factor $\gamma$
	%		\Ensure 
			\State {Initialize a policy network with random weight $\theta_i^{(0)}$ and an intrinsic reward network with random weight $\eta_i$}.
			\Repeat
			\For {$k=1,2,\dots,N_\text{ep}$}
			\parState {Interact with the environment for one episode using $\pi_{\theta_i^{(k-1)}}$. }
			\parState {Store the experience within the episode rollout $T_{\text{E},i}^{(k)}$ (\ref{ep_rollout}) and the lifetime rollout $T_{\text{L},i}$ (\ref{life_rollout})}
			\parState {Update the policy parameter $\theta_i^{(k-1)}$ by exploiting $T_{\text{E},i}^{(k)}$ as described in Section \ref{policy_update}}
			\EndFor
			\parState {Update the intrinsic reward network with parameter $\eta_i$ by exploiting  $T_{\text{L},i}$ as described in Section \ref{intrinsic_update}.}
			\parState {Set $\theta_i^{(0)}\leftarrow \theta_i^{(N_\text{ep})}$.} 
			\Until the intrinsic reward network converges.
		\end{algorithmic}
	}
\end{algorithm}

At each episode $k$, each agent $i$ generates an experience 
interacting with the environment for $T_{\text{ep}}$ time steps using its policy and its intrinsic reward network. In detail, at each time step $t$, the experience of agent $i$ is stored inside the episode rollout
 \begin{equation}
 \label{ep_rollout}
      T_{\text{E},i}^{(k)}= \left\{\left(o^t_i,a^t_i, \pi_{\theta^{(k-1)}_i} \left (a^t_i \mid o^t_i \right), R_{\text{ext}}^{t+1},  R_{\text{in},\eta_i}^{t+1}\right)\right\}_{t=0}^{T_{\text{ep}}-1},
 \end{equation}
 and the lifetime rollout
\begin{equation}
 \label{life_rollout}
T_{\text{L},i}= \left\{T_{\text{E},i}^{(k)}\right\}_{k=1}^{N_\text{ep}}.
\end{equation} 
Episode-by-episode, each agent $i$ updates its policy parameter $\theta_i^{(k-1)}$ following the procedure is described in Section \ref{policy_update}. 
At the end of a lifetime, each agent updates the intrinsic reward network parameter $\eta_i$ following the procedure described in Section \ref{intrinsic_update}. The overall procedure is carried out until the intrinsic reward network reaches convergence.

%After that, each agent updates the intrinsic reward network parameter $\eta_i$  in the direction of the gradient of the expected lifetime extrinsic return, by exploiting  $T_{\text{L},i}$, as described in Section \ref{intrinsic_update} (see Fig. \ref{training_periodicity}). The overall procedure is carried out until the intrinsic reward network reaches convergence.
%Episode-by-episode, each agent $i$ updates its policy parameter $\theta_i^{(k-1)}$ by exploiting $T_{\text{E},i}^{(k)}$, in the direction of the gradient of expected lifetime overall return, which considers both the intrinsic and extrinsic rewards. The procedure is described in Section \ref{policy_update} and is carried out for a lifetime, composed of $N_\text{ep}$ episode (see Fig. \ref{training_periodicity}).
%After that, each agent updates the intrinsic reward network parameter $\eta_i$  in the direction of the gradient of the expected lifetime extrinsic return, by exploiting  $T_{\text{L},i}$, as described in Section \ref{intrinsic_update} (see Fig. \ref{training_periodicity}). The overall procedure is carried out until the intrinsic reward network reaches convergence.

\subsection{Updating the Policy Parameter $\theta_i$}
\label{policy_update}
In this subsection, we describe how to update the policy parameter of each agent $i$. Specifically, at the end of episode $k$, the update of $\theta^{(k-1)}_i$ is performed so as to maximize the expected episodic cumulative overall return of episode $k-1$
\begin{equation}
    J_{\text{ep},\text{ov}, i}^{(k-1)}=  \mathbb{E}_{o_i^0,a_i^0,\dots o_i^{T_\text{ep}-1},a_i^{T_\text{ep}-1}} \left [G_{\text{ep},\text{ov},i}^{(k-1)}  \right ],
\end{equation}
where $a_i^t\sim\pi_{\theta^{(k-1)}_i}\left(a_i^t\mid o_i^t\right)$.
This update can be done by using a simple policy gradient method, as follows
\begin{equation}
\begin{split}
\theta^{(k)}_i &= \theta^{(k-1)}_i + \alpha \nabla_{\theta^{(k-1)}_i} J_{\text{ep},\text{ov}, i}^{(k-1)}.
\end{split}
\end{equation}
The policy gradient theorem \cite{NIPS1999_464d828b} shows that, given the episode rollout $T_{\text{E},i}^{(k)}$, the update can be computed as\footnote{Other approaches exploiting other policy gradient methods (REINFORCE \cite{Reinforce_1}, PPO \cite{PPO_1}, TRPO \cite{TRPO_1}) can be used.}
\begin{equation}
\label{pol_update}
    \theta^{(k)}_i  \approx  \theta^{(k-1)}_i + \alpha G_{\text{ep},\text{ov},i}^{(k-1)}\nabla_{\theta^{(k-1)}_i} \log{\pi_{\theta^{(k-1)}_i}} \left (a_i^t|o_i^t \right ).
\end{equation}
% \end{split}
% \end{equation}
%, the inputs $\{o_i^t\}_{t=0}^{T_\text{ep}-1}$ and $\{a_i^t\}_{{t=0}}^{T_\text{ep}-1}$ are extracted and provided as input to the intrinsic reward network. 
%After, taking into account $\{R_\text{ext}^{t+1}\}_{t=0}^{T_\text{ep}-1}$ and the computed intrinsic reward values,
%Then, we calculate, for each agent $i$, the expected episodic cumulative overall return as 
%\begin{equation}
%J_{\text{ep},\text{ov}, \text{i}}(\pi_i) = 
%G_{\text{ep},\text{ov}, \text{i}}.
% \mathbb{E}_{{o_\text{i}^0},{a_\text{i}^0},R_\text{ext}^{1},  \dots, {o_\text{i}}^{T_\text{ep}-1},{a_\text{i}^{T_\text{ep}-1}}, {R_\text{ext}^{T_\text{ep}-1}}} \left [
% G_{\text{ep},\text{ov}, \text{i}}  \right ].
%\end{equation}
%Finally, the update of $\theta_i$ can be performed following using regular policy gradient
%sing the overall rewards as the reward:
% \begin{equation}
% \begin{split}
%     \theta' &= \theta + \alpha \nabla J^\text{ex+in}(\theta) \\
%     &\approx \theta + \alpha G^\text{ex+in}(s_t,a_t) \nabla_\theta \log_{\pi_\theta}(a_t|s_t)
% \end{split}
% \end{equation}

\subsection{Updating the Intrinsic Reward Parameter ($\eta_i$)}
\label{intrinsic_update}
Given a lifetime and the updated policy parameters at the end of the lifetime $\left ({\theta}_i^{\left(N_\text{ep}\right)}\right )$, we update the intrinsic reward network parameter for each agent $i$ with the aim of maximizing the expected lifetime extrinsic return
\begin{equation}
    J_{\text{life},\text{ext}}=\mathbb{E}_{o_i^0,a_i^0,\dots o_i^{T-1},a_i^{T-1}} \left [G_{\text{life},\text{ext}}  \right ],
\end{equation}
where $a_i^t\sim\pi_{\theta^{(N_\text{ep})}_i}\left(a_i^t\mid o_i^t\right)$.
Similarly as the policy parameters update, this update can be done by using a simple policy gradient method, as follows
\begin{equation}
\label{intr_update}
\eta_i' = \eta_i + \beta \nabla_{\eta_i} J_{\text{life},\text{ext}}.
\end{equation}
Intuitively, updating $\eta_i$ requires estimating the effect such a change would have on the extrinsic value through the change in the policy parameters. To obtain this, we compute the meta-gradient $\nabla_{\eta_i} J_{\text{ext},\text{life} }$ exploiting the chain rule as follows:
\begin{equation}
\label{nabla_intr}
 \nabla_{\eta_i} J_{\text{life},\text{ext} } = \nabla_{{\theta}_i^{\left(N_\text{ep}\right)}}  J_{\text{life},\text{ext}} \nabla_{\eta_i} {\theta}_i^{\left(N_\text{ep}\right)}.
\end{equation}

Moreover, the first gradient can be approximated by means of the policy gradient theorem \cite{NIPS1999_464d828b} as
\begin{equation}
   \nabla_{{\theta}_i^{\left(N_\text{ep}\right)}}  J_{\text{life},\text{ext}}  \approx  G_{\text{life},\text{ext}} \nabla_{\theta^{(N_\text{ep})}_i}\log{\pi_{\theta^{(N_\text{ep})}_i}} \left (a_i^t|o_i^t \right ).
\end{equation}
%The update should consider the influence that ${\theta}_i^{\left(N_{ep}\right)}$ has on $J_{\text{life},\text{ext}}$, being ${\theta}_i^{\left(N_{ep}\right)}$ also updated on the basis of $\eta_i$. 

We note that  a new lifetime should be computed with the updated policy parameters $\theta^{(N_\text{ep})}_i$ to calculate this gradient.
For avoiding this, we reuse the lifetime generated by the original policy parameters $\theta^{(k)}_i$, with $k=1,\dots,N_\text{ep}$, by means of the \emph{importance sampling ratio} \cite{https://doi.org/10.48550/arxiv.2207.06131}. Hence, we exploit the lifetime rollout $T_{\text{L},i}$, and rewrite the gradient computation as follows:
\begin{equation}
\label{appr_intr}
   \nabla_{{\theta}_i^{\left(N_\text{ep}\right)}}  J_{\text{life},\text{ext}}  \approx  G_{\text{life},\text{ext}} \frac{\nabla_{\theta^{(N_\text{ep})}_i}{\pi_{\theta^{(N_\text{ep})}_i}} \left (a_i^t|o_i^t \right )}{\pi_{\theta^{(k)}_i} \left (a_i^t|o_i^t \right )}.
\end{equation}

\begin{figure*}[!t]
	\centering
	\subfloat[$P=1$.]{\includegraphics[height=1.9in]{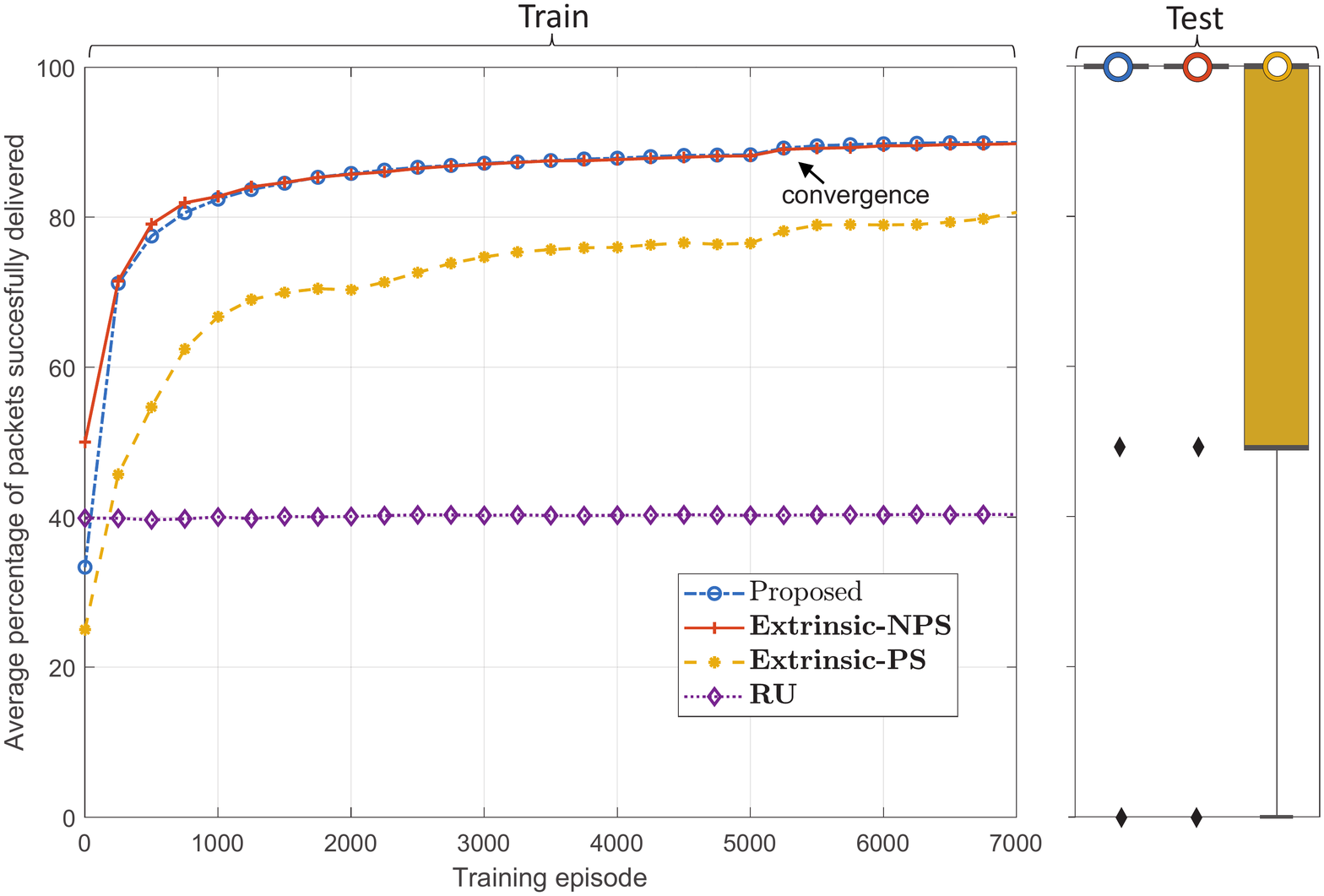}%
		\label{convergence_curves_P_1}
	}
	\hfil
	\subfloat[$P=2$.]
	{\includegraphics[height=1.9in]{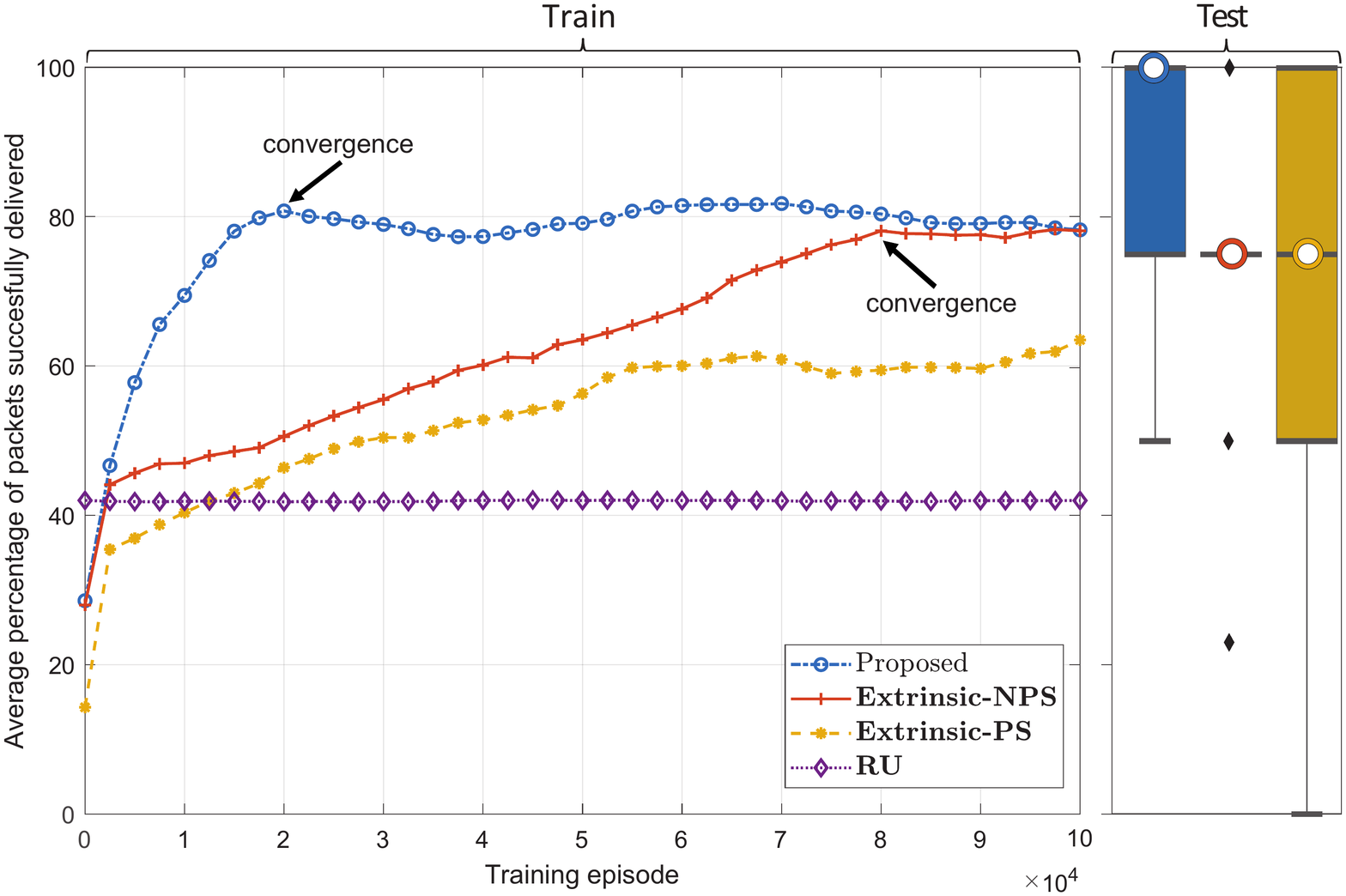}%
		\label{convergence_curves_P_2}}
	\caption{Convergence curves of all the considered algorithms varying the  number of packets to deliver and testing results. In the boxplots,  the  circle is the median, the colored box represents the interquartile range (IQR) from the first quartile (Q1) to the third quartile (Q3), the vertical lines represent the minimum (Q1-1.5 IQR) and the maximum  (Q3+1.5 IQR) values, the diamonds are outliers.}
	\label{convergence_curves}
\end{figure*}

\section{Simulation Results and Analysis}
\label{sim_and_ana}
In this section, we  examine the convergence performance of the proposed learning framework in terms of percentage of  successfully delivered dPDUs vs. the number of training episodes. %For our simulations,  we considered a scenario with two UEs and one BS. 
The list of the simulation parameters is reported in Table \ref{simulation_parameters}.

\begin{table}[!ht]
	\renewcommand{\arraystretch}{1}
	\caption{Training algorithm Parameters}
	\label{simulation_parameters}
	\centering
		\begin{threeparttable}
	\begin{tabular}{l c c}
		\hline
		\multicolumn{1}{c}{\textbf{Parameter}} & 
	    \multicolumn{1}{c}{\textbf{Symbol}} & 
    	\multicolumn{1}{c}{\textbf{Value}}\\
		\hline
		Number of UEs & $N$ & 2\\
		Discount factor  & $\gamma$ & 0.99 \\
	%	Epsilon value & $\epsilon$ & 0.1 \\
	    Memory length  & $M$ & $3$ \\
    	Duration of episode & $T_{\text{ep}}$ &	32\\
    	Number of episodes per lifetime & $N_\text{ep}$& 250 \\
    	Number of dPDUs to deliver& $P$ & $\{1, 2\}$\\
    	Balancing parameter& $\lambda$ & 1\\
    	Act. function, intrinsic reward network &&\{i, t\}\tnote{a}\\%\footnotemark[\ref{note1}]\\
    	Learning rate, intrinsic reward network&  $\beta$& $7\cdot 10^{-4}$ \\
    	%Number of test episodes & %$N_{\text{test}}$  & 1000\\
    	Neurons per layer, intrinsic reward network & & 128\\
    	Act. function per layer, policy network &&\{t, t, s\}\tnote{a}\\%\footnotemark[\ref{note1}]\\
    	Learning rate, policy network&  $\alpha$& $3\cdot 10^{-4}$ \\
    	Neurons per layer, policy network & & 64\\
	 	%Reward function parameter & $\rho$& 3\\
		\hline
% 		\multicolumn{1}{c}{\textbf{$M_{\mathcal{O}}$ and $M_{\mathcal{O_{\phi}}}$   Parameter} } & 
% 	    \multicolumn{1}{c}{\textbf{Symbol}} & 
%     	\multicolumn{1}{c}{\textbf{Value}}\\
% 		\hline	
% 		Num. of neurons per hidden layer, evaluator & & 64\\
% 	%	\hline
% %		\hline
% 		Num. of neurons per hidden layer, actor & & 64\\
% 	%	\hline
% 		Memory length  & $M$ & $1$ \\
% 	%	\hline		
% 		Learning rate &  $lr_{M}$& $10^{-3}$ \\
% 	%	\hline
% 		Number of training episodes & $N_{\text{tr}}$& $20$k\\
% 	%	\hline
% 		Act. function per layer, evaluator &&\{t, t, i\}\myfootnotemark{}\\
% 	%	\hline
% 		Act. function per layer, actor &&\{t, t, s\}\myfootnotemark{}\\
% 	%	\hline				
% 		Clipping value  & $\psi$ & $0.2$ \\
% 		\hline
	\end{tabular}
	\begin{tablenotes}\footnotesize
		\item [a] i = identity function, t = tanh function, s = softmax function.
	\end{tablenotes}
\end{threeparttable}
\end{table}
%\myfootnotetext{i = identity function, t = tanh function, s = softmax function\label{note1}}

The results of the proposed method are compared against the following baselines. 
\begin{itemize}
    \item \textbf{Extrinsic-NPS}: An independent policy is trained for each agent $i$ to maximize the expected episodic cumulative extrinsic return ($J_{\text{ep},\text{ext}}$). 
    \item \textbf{Extrinsic-PS}: A shared policy is trained among all the agents to maximize $J_{\text{ep},\text{ext}}$, as in \cite{nokia_paper_1}.
    \item \textbf{Random Uniform (RU)}: Regardless of the observation $o_i^t$, each agent $i$ select its action uniformly from $\mathcal{A}$.
\end{itemize}
%All statistical results are averaged over a large number of independent runs. 
Fig. \ref{convergence_curves} plots the  percentage of successfully delivered packets vs. training episodes with respect to baselines averaged over 10 independent training sessions.  After assessing the training phase, we select the best trained instance for each solution in terms of average percentage of successfully delivered packets. Then, we test them in 1000 episodes and show the related statistics by using boxplots.
%Then, we show their results in the testing phase using boxplots.
% Each figure differs from the other in terms of lifetime duration.
% In particular, the lifetime duration impacts significantly the training performance of the proposed method. In fact, this value is crucial to balance the policy network and the intrinsic reward network updates. If the intrinsic reward network is updated too fast, then the updates of the policy network is not stable due to the intrinsic reward values variance.
% Conversely, if the lifetime duration is too long then the convergence time is long, and there is no benefit from the introduction of the intrinsic reward network.

Specifically, Fig. \ref{convergence_curves_P_1} shows the simulation results in the case of $P = 1$ packet to deliver. We observe that the proposed algorithm and $\textbf{Etrinsic-NPS}$ require $5.1\cdot 10^3$ iterations to reach convergence. Conversely, $\textbf{Etrinsic-PS}$ has not converged within $7\cdot 10^3$ episodes.
This shows that the introduction of additional features (intrinsic reward and lifetime update), does not introduce any significant improvement in the case of a simple transmission scenario.
% the convergence performance with lifetime duration equal to 10 episodes. Therein, we observe that the proposed method achieves rapidly ($\approx 15 \cdot 10^3 $ episodes) a maximum in terms of performances ($82.5\%$) and after that the performance slightly decreases. Conversely, the other methods start slower but they continue to increase their performances reaching a plateau at around $100 \cdot 10^3 $ episodes, with a maximum value of performance equal to $88\%$ for the \textbf{Extrinsic-NPS}, and $67\%$ for the \textbf{Extrinsic-PS}. Obviously, the $\textbf{RU}$ curve does not improve during the training and the performances are constant to $40\%$. In Fig. \ref{convergence_curves}b, we show the performance with lifetime duration equal to 250 episodes. Since the baseline methods do not depend on the lifetime duration, they show the same performance. Conversely, the proposed method with a lifetime duration of 250 episodes exhibits a significant improvement both in terms of transmission performance ($89\%$) and convergence speed ($\approx 2\cdot 10^3$ episodes).
In Fig. \ref{convergence_curves_P_2} we show the performances when  $P = 2$  packets need to be delivered. Therein, the proposed method reaches convergence in almost $2\cdot 10^4$ training episodes, which is $75\%$ less than the number of episodes required for the \textbf{Extrinsic-NPS} method. This is because, in this more complex scenario, the proposed method provides additional information with the correct periodicity to the policy update process. Fig. \ref{convergence_curves_P_2} shows also that the proposed method achieves a maximum service success rate of $81\%$, that is $4\%$ better than the maximum performances of the \textbf{Extrinsic-NPS} one. As regards the other baseline, \textbf{Extrinsic-PS} does not reach  convergence within the considered training interval.
As concerns the testing phase, our solution exhibits an interquartile range between $75\%$ and $100\%$ of packet successfully delivered, which is the best result, as show in Fig. \ref{convergence_curves_P_2}.

Summarizing, the proposed method yields better convergence speed with better transmission performances in the case of a complex scenario in which the additional information is key for policy parameter tuning. 
%Conversely, when the scenario is too simple the performances are not enhanced and remain the same of the most performing baseline method.

\section{Conclusions}
\label{conclusion}
We have proposed a novel multi-agent reinforcement learning framework for MAC protocol learning, which in addition to using the classical extrinsic team reward, learns an individual intrinsic reward for each agent based on its history. Each agent uses two modules, namely a policy network and an intrinsic reward network. These two modules are updated with a different periodicity to obtain better learning results in terms of convergence speed. Specifically, the policy network is trained within an episode, while the intrinsic reward network is trained over a fixed number of subsequent episodes, called lifetime. We formulated an optimization problem that seeks to maximize the number of successfully transmitted packets.
Our results demonstrate that exploiting these two modules with two different learning periodicities induces a faster convergence speed compared to several baseline solutions. 

\section*{Acknowledgments}
This work was partially supported by the European Union under the Italian National Recovery and Resilience Plan (NRRP) of NextGenerationEU, partnership on ``Telecommunications of the Future'' (PE00000001 - program ``RESTART''), by the Italian MUR PON 2014-2020 under Project
``reCITY - Resilient City - Everyday Revolution'' (cod. ARS01\textunderscore00592, CUP B69C21000390005), and by the European Union’s Horizon Europe program through the project CENTRIC.
%by CHIST-ERA CONNECT project, by Horizon EU-CENTRIC project, and by FutureWei gift funding. 
\bibliographystyle{IEEEtran}
\bibliography{IEEEabrv,Bibliografia}
\end{document}